\documentclass[prb,twocolumn,showpacs,preprintnumbers,amsmath,amssymb]{revtex4}

\usepackage{graphicx}
\usepackage{dcolumn}
\usepackage{bm}

\begin{document}

\title{Quantum phase transitions in the one-dimensional asymmetric Hubbard model:
a bosonization study}
\author{Z. G. Wang$^{1,2}$}
\author{Y. G. Chen$^{1}$}
\author{S. J. Gu$^{2}$} \altaffiliation{Email: sjgu@phy.cuhk.edu.hk
\\URL: http://www.phystar.net/}
\address{$^{1}$ Department of Physics, Tongji University, Shanghai, 200092, China}
\address{$^{2}$ Department of Physics and the Institute of Theoretical Physics,
the Chinese University of Hong Kong, Hong Kong, China}
\begin{abstract}
The quantum phase transitions in the one-dimensional asymmetric Hubbard model
are investigated with the bosonization approach. The critical conditions for the
transition from density wave to phase separation, the correlation functions and
their exponents are obtained analytically. Our results show that the difference
between the hopping integrals for up- and down-spin electrons is crucial for the
happening of the phase separation. When the difference is large enough, the
phase separation will appear even if the on-site interaction is small.
\end{abstract}



\date{\today}
\maketitle

\section{Introduction}

The Hubbard model (HM)\cite{Hubbardmodel} is one of the simplest nontrivial
models of interacting spin-1/2 electrons on a lattice. Its Hamiltonian reads
\begin{equation}
H_{\rm Hub}=-t\sum_{j=1}^L\sum_{\delta=\pm 1} c^\dagger_{j,\sigma}c_{j+\delta,
\sigma}+U\sum_{j}n_{j\uparrow }n_{j\downarrow },
\end{equation}
where $c^\dagger_{j,\sigma}$ and $c_{j,\sigma},\sigma=\uparrow,\downarrow$ are
creation and annihilation operators for electrons with spin $\sigma$ at site $j$
respectively, $n_\sigma=c_\sigma^\dagger c_\sigma$, $t$ is the hoping integral,
and $U$ denotes the strength of on-site interaction. In one dimension (1D) the
HM can be solved exactly by the Bethe-ansatz method.\cite{E. Lieb,FHLEsslerb}
The wave function and the energy of $N=N_{\uparrow }+N_{\downarrow }$ electrons
on a chain with $L$ sites can be written in terms of $N$ pseudo-momentum
variables and $N_{\downarrow }$ spin rapidities. Although the energy spectra
have been known for many years, the calculation of the correlation functions
proved to be a delicate problem. \cite{JCarmelo88,M.Ogata,A.Parola,Frahm90} The
numerical evaluations of the correlation functions \cite{M.Ogata} and the
analytic results \cite{A.Parola} indicated clearly that the 1D HM is a
Tomonaga-Luttinger liquid (TLL).\cite{S. Tomonaga,F.D.M. haldane} Assuming that
the 1D HM is TLL, it then becomes possible to calculate the correlation
functions from the knowledge of the energy spectra \cite{J.Solyom}. Using this
procedure Schulz \cite{H.J.Schulz} studied the correlation-function exponents
for different $U$ and band filling $n$. It is also shown\cite{K. Penc} that the
large scale behavior of the spin and charge degrees of freedom can be described
by two decoupled boson field theories with dynamics governed by the TLL
Hamiltonian in the small and large $U$ regimes.

Another nontrivial model is the Falicov-Kimball model \cite{L. M. Falicov} which
consists of localized ions and itinerant spinless fermions. The Hamiltonian of
the Falicov-Kimball model reads
\begin{equation}
H_{\rm FK}=-t\sum_{j=1}^L \sum_{\delta=\pm 1}
c^\dagger_{j}c_{j+\delta}+U\sum_{j}n_{j}w_{j},
\end{equation}
where $c^\dagger_j$ are creation operators for spinless fermions, and the
configuration $\{w_j\}$ denote spatial distribution of ions. Clearly, the FKM
can be viewed as a modification of the HM in the sense that the one kind of
fermions, such as down-spin fermions, has infinite mass, and hence does not
move. Nevertheless, the physics of the FKM is completely different. In the
neutral case where each particle concentration equals $1/2$, it was \cite{T.
Kennedy} proved that the system always orders in an alternating ``chessboard''
phase at a finite transition temperature in all dimensions greater than 1. This
ordered phase can be interpreted as the transition from a high-temperature
homogeneous (liquid/gas) phase to a low temperature ordered (solid) phase.
Freericks\cite{J. K. Freericks} showed that the model (on a hypercubic lattice)
also displayed incommensurate order, segregation or phase separation. The 1D
case of the FKM has also been extensively studied. Since there is no
finite-temperature phase transition, the system can have phase transition in the
ground state. The numerical solutions \cite{J. K. Freericks1} produced a
conjecture for the case $n _{e}+n _{i} < 1$ with $n_{e}=N_e/N, n_i=N_i/N$ and
the screened Coulomb interaction $U$ is large enough, the system will segregate
into an empty lattice (with no ions and all the electrons), and a full lattice
(with all the ions and no electrons). This conjecture was later proven to be
true by Lemberger \cite{P. Lemberger}. For any dimensional HFM, Freericks {\it
etal} \cite{JKFreericks02} gave a theorem that the strong correlation can leads
to PS.

The relation between the HM and the FKM is straightforward. In order to have a
unified framework, the asymmetric Hubbard model (AHM) has been introduced
naturally. \cite{UBrandt91,GFath95,CAMacedo02,DUeltschi04} Its Hamiltonian reads
\begin{eqnarray}
H_{\rm AHM}=-\sum_{j=1}^{L}\sum_{\delta=\pm 1} \sum_\sigma t_\sigma
c^\dagger_{j,\sigma}c_{j+\delta, \sigma}+U \sum_{j=1}^L n_{j, \uparrow}n_{j,
\downarrow}, \label{eq:Hamiltonian_AHM}
\end{eqnarray}
where $t_\sigma$ is $\sigma$-dependent hoping integral. Clearly, if $t_\uparrow
=t_\downarrow$, the AHM becomes the HM, and if $t_\downarrow =0 $, it becomes
the FKM. The Hamiltonian (\ref{eq:Hamiltonian_AHM}) has U(1)$\otimes$U(1)
symmetry for general $t_\sigma$, and the electron number $N_\downarrow,
N_\uparrow$ are conserved respectively. In the condensed matter physics, the AHM
is believed to describe many physical phenomena, such as superconductor, valence
fluctuating, and heavy fermions.\cite{VJEmery87,CMVarma85} In the recent
development of the optical lattice, it has been pointed out that the AHM can be
used to describe a mixture of two species of fermionic atoms in optical
lattices. \cite{MACazalilla05,SJGu05}

According to the fact that the ground states of the Hamiltonian
(\ref{eq:Hamiltonian_AHM}) in its two limiting cases: the HM
($t_\uparrow=t_\downarrow$) and the FKM ($t_\downarrow=0$) belong to two
different universality classes, a quantum phase transition was suggested to
happen in the phase diagram defined in the $U-t_\downarrow$
plane.\cite{GFath95,DUeltschi04,MACazalilla05} Nevertheless, the quantitative
phase diagram for the case away from half-filling had never been obtained until
a recent work by Gu, {\it et al} \cite{SJGu05}. In their work, the quantum
entanglement \cite{SJGuPRL} between a local part and the rest of the system, and
the structure factor of charge-density-wave (CDW) for down-spin electrons are
used to identify the critical point. Here, we are going to study the
ground-state phase diagram of the AHM away from the half filling with the
strategy of bosonization method. \cite{bosonization_rev} Differ from the
numerical approach \cite{SJGu05} which captures the physics from the finite-size
analysis for small systems, our work aim to give definitely analytical results
from the point of view of field theory. The paper is organized as follows. In
section \ref{sec:bos}, we derive the bosonized form of the 1D AHM, and clarify
the role of the some terms in the Hamiltonian. In section \ref{sec:res}, we
first diagonalize the effective Hamiltonian in which some irrelevant terms are
ignored, then obtain the instability conditions for the phase separation and
compare them with the numerical results of a finite sample. We also obtain the
analytical expressions for the correlation functions of charge-density-wave,
spin-density-wave (SDW), singlet-superconductivity (SS),
triplet-superconductivity (TS) fluctuations, as well as the corresponding
exponents. Finally, a brief summary and acknowledgement are given in section
\ref{sec:sum}.

\section{The Bosonized Form of one-dimensional asymmetric Hubbard model}
\label{sec:bos}

The convenient way to analyze the 1D AHM is to bosonize the Fermi operators and
convert them to a quantum theory of two Bose fields \cite{bosonization_rev,D.
Shelton}. In the framework of the standard bosonization method the HM are
expressed in terms of canonical Bose fields and their dual counterparts as,
\begin{eqnarray}
H_{B} &=&\frac{v_{c}}{2}\int dx\left[ \frac{1}{K_{c}}\left( \partial
_{x}\phi _{c}\right) ^{2}+K_{c}\pi _{c}^{2}\right]  \\
&&+\frac{v_{s}}{2}\int dx\left[ \frac{1}{K_{s}}\left( \partial
_{x}\phi
_{s}\right) ^{2}+K_{s}\pi _{s}^{2}\right]\nonumber \\
&&+\delta v\int dx\left[ \pi _{c}\pi _{s}+\partial _{x}\phi _{c}\partial
_{x}\phi _{s}\right] \nonumber \\
&&+\frac{U}{2\pi ^{2}a}\int dx\cos \left( \sqrt{8\pi }\phi _{s}\right) \nonumber \\
&&+\frac{U}{2\pi ^{2}a}\int dx\cos \left( \sqrt{8\pi }\phi
_{c}+2(k_{F\uparrow}+k_{F\uparrow})x\right), \nonumber
\end{eqnarray}
with
\begin{widetext}
\begin{eqnarray} v_{c} &=&a\sqrt{{t_{\uparrow }\sin (k_{F\uparrow
}a)+t_{\downarrow }\sin (k_{F\downarrow }a)}\left( {t_{\uparrow }\sin
(k_{F\uparrow
}a)+t_{\downarrow }\sin (k_{F\downarrow }a)}+\frac{U}{2\pi }\right) }, \\
\frac{1}{K_{c}}&=&\sqrt{1+\frac{U}{2\pi \left[ t_{\uparrow }\sin
(k_{F\uparrow
}a)+t_{\downarrow }\sin (k_{F\downarrow }a)\right] }}, \\
v_{s} &=&a\sqrt{{t_{\uparrow }\sin (k_{F\uparrow }a)+t_{\downarrow
}\sin (k_{F\downarrow }a)}\left( {t_{\uparrow }\sin (k_{F\uparrow
}a)+t_{\downarrow }\sin (k_{F\downarrow }a)}-\frac{U}{2\pi }\right) }, \\
\frac{1}{K_{s}} &=&\sqrt{1-\frac{U}{2\pi \left[ t_{\uparrow }\sin
(k_{F\uparrow}a)+t_{\downarrow }\sin (k_{F\downarrow }a)\right] }}, \\
\delta v &=&{a}\left[ t_{\uparrow }\sin (k_{F\uparrow }a)-t_{\downarrow }\sin
(k_{F\downarrow }a)\right].
\end{eqnarray}
\end{widetext}
Here, the Bose field $\phi _{c},\phi _{s}$ present the charge and spin degrees
of freedom, respectively. $k_{F\uparrow }$ and $k_{F\downarrow }$ are the Fermi
momentum for up- and down-spin electrons,  $k_{F\uparrow }={\pi n_\uparrow}/{a},
k_{F\downarrow }={\pi n_\downarrow}/{a},$ with $n_\uparrow=N_\uparrow/L$ and
$n_\downarrow=N_\downarrow/L$ are the filling densities for up- and down-spin
electrons, respectively; and $a$ lattice constant. $v_{c,s}$ are the propagation
velocities of the charge and spin collective modes of the decoupled model
($\delta v$ = 0), and $K_{c,s}$ are the stiffness constants.

It can be seen that this model is the standard HM at unpolarized case with both
Bose fields decoupling. And if model is at half-filling band, $k_{F\uparrow
}+k_{F\downarrow }={\pi }/{a},$ the Umklapp term (the last term in $H_B$) is
important and the HM is in the SDW phase with $U>0$ or in the CDW phase with
$U<0$. As the model shifts away from the half-filling band, the Umklapp
interaction can be cancelled. The difference of hopping integral and the filling
densities of up- and down-spin electrons in the system appear as an effect that
breaks the spin-charge separation as reveals the presence of the third term in
the last equation.

Here, we discuss the quantum transition in the case of away from the half
filling which means that the Umklapp interaction can be ignored. In general
$t_{\uparrow },t_{\downarrow }>0$. In the parameter region of $U>0$, since
$K_{s}>1$, so the term of $\cos \left( \sqrt{8\pi }\phi _{s}\right) $ is
irrelevant whenever the system is at unpolarized case, in the one-loop
approximation we can cancel it directly and get
\begin{eqnarray}
H_{\rm Beff} &=&\frac{v_{c}}{2}\int dx\left[ \frac{1}{K_{c}}\left(
\partial _{x}\phi _{c}\right) ^{2}+K_{c}\pi _{c}^{2}\right]  \nonumber \\
&&+\frac{v_{s}}{2}\int dx\left[ \frac{1}{K_{s}}\left( \partial _{x}\phi
_{s}\right) ^{2}+K_{s}\pi _{s}^{2}\right]  \nonumber \\
&&+\delta v\int dx\left[ \pi _{c}\pi _{s}+\partial _{x}\phi _{c}\partial
_{x}\phi _{s}\right].  \label{equ9}
\end{eqnarray}

\section{Results and discussions}
\label{sec:res}

The Hamiltonian (\ref{equ9}) can be diagonalized in terms of two new phase
fields which contains a mixture of spin and charge degrees of freedom. The
propagation velocities of these collective modes are

\begin{widetext}
\begin{equation}
v_{\pm }^{2}=\frac{v_{c}^{2}+v_{s}^{2}}{2}+\delta v^{2}\pm \sqrt{\left(
\frac{v_{c}^{2}-v_{s}^{2}}{2}\right) ^{2}+\delta v^{2}\left[
v_{c}^{2}+v_{s}^{2}+v_{c}v_{s}\left( K_{c}K_{s}+\frac{1}{K_{c}K_{s}}\right)
\right]. }
\end{equation}
As $\delta v\rightarrow 0,v_{+}\rightarrow \max \left( v_{c},v_{s}\right) $
and $v_{-}\rightarrow \min \left( v_{c},v_{s}\right) $. As $\delta v$
increases, $v_{-}$ decreases until vanishes at the points
\begin{eqnarray}
\delta v_{1}^{2} &=&v_{c}v_{s}\frac{1}{K_{c}K_{s}} \\
\delta v_{2}^{2} &=&v_{c}v_{s}K_{c}K_{s}
\end{eqnarray}%
At these points, the freezing of the lower bosonic mode is accompanied by a
divergence in the charge and spin response functions. The static charge
compressibility $\kappa $ diverges at $\delta v$ =$\delta v_{1}$, or $%
\delta v$ =$\delta v_{2}$ . it behaves as

\begin{equation}
\kappa =\kappa _{0}\left[ 1-\frac{\delta v}{\delta v_{1(2)}}\right] ^{-1},%
\text{ \ \ \ \ }\kappa _{0}=\frac{2K_{c}}{\pi v_{c}},
\end{equation}%
Beyond these points the susceptibilities becomes negatives. This behavior of the
static response functions together with the vanishing of the collective modes
velocity indicates that the system becomes unstable\cite{J. Voit} and undergoes
a first order phase transition\cite{J. Voit1}. This instability is known as
phase separation and has been shown to occur in the extended
HM\cite{HQLin00,KPenc94} and in the $t-J$ model\cite{VJEmery90,MOgata91}.

In our case, we obtain

\begin{eqnarray}
\delta v_{1}^{{}} &=&\sqrt{v_{c}v_{s}K_{c}K_{s}}=a{t_{\uparrow
}\sin (k_{F\uparrow }a)+t_{\downarrow }\sin (k_{F\downarrow }a)}, \\
\delta v_{2} &=&\sqrt{\frac{v_{c}v_{s}}{K_{c}K_{s}}}=a{t_{\uparrow
}\sin (k_{F\uparrow }a)+t_{\downarrow }\sin (k_{F\downarrow
}a)}\sqrt{1-\left( \frac{U}{2\pi\left( t_{\uparrow }\sin
(k_{F\uparrow }a)+t_{\downarrow }\sin (k_{F\downarrow
}a)\right)}\right) ^{2}}.
\end{eqnarray}%
It is obvious that $\delta v_{1}\geq \delta v_{2},$ so the system is
in PS phase state as
\begin{equation}
\delta v\geq \delta v_{2},
\end{equation}
i.e.,
\begin{equation}
\frac{t_{\uparrow }\sin (k_{F\uparrow }a)-t_{\downarrow }\sin
(k_{F\downarrow }a)}{2}\geq \frac{t_{\uparrow }\sin (k_{F\uparrow
}a)+t_{\downarrow }\sin (k_{F\downarrow }a)}{2}\sqrt{1-\left( \frac{U}{%
 2 \pi \left( t_{\uparrow }\sin (k_{F\uparrow }a)+t_{\downarrow }\sin
(k_{F\downarrow }a)\right)}\right) ^{2}}.  \label{equ11}
\end{equation}
\end{widetext}
Then we obtain the PS phase state condition
\begin{equation}
\left({t_{\downarrow }\sin (k_{F\downarrow
}a)}\right)\left({t_{\uparrow }\sin (k_{F\uparrow }a)}\right)\leq
\left( \frac{U}{4\pi}\right) ^{2}. \label{equ12}
\end{equation}

Let us now focus our attention on the correlation functions. Our interest in
this work is observing the algebraic decay of the instantaneous correlation
functions at zero temperature and studying how the exponents get modified from
the standard HM. The operators for CDW, SDW, SS, and TS fluctuations in their
bosonized form are\cite{J.Solyom}
\begin{eqnarray}
&&O_{\rm CDW}^{+}(x)=\Psi _{1\uparrow }^{+}\Psi _{2\uparrow } \nonumber \\
&&\;\; =\frac{1}{2\pi \epsilon}\exp\left[\sqrt{2\pi }i(\phi _{\rho }+\phi
_{\sigma })+2ik_{F}x\right], \nonumber
\\
&&O_{\rm SDW}^{+}(x)=\Psi _{1\uparrow }^{+}\Psi _{2\downarrow } \nonumber \\
&&\;\; =\frac{1}{2\pi \epsilon}\exp\left[\sqrt{2\pi }i\left(\phi _{\rho
}(x)-\int_{-\infty }^{x}dy\Pi _{\sigma
}(y)\right)-2ik_{F}x\right],  \nonumber \\
&&O_{\rm SS}^{+}(x)=\Psi _{1\uparrow }^{+}\Psi _{2\downarrow }^{+} \nonumber \\
&&\;\; =\frac{1}{2\pi \epsilon }\exp\left[\sqrt{2\pi }i\left(-\int_{-\infty
}^{x}dy\Pi _{\rho }(y)+\phi
_{\sigma }(x)\right)\right],  \nonumber \\
&&O_{\rm TS}^{+}(x)=\Psi _{1\uparrow }^{+}\Psi _{2\uparrow }^{+}  \\ \nonumber
&&\;\; =\frac{1}{2\pi \epsilon }\exp\left[\sqrt{2\pi }i\left(-\int_{-\infty
}^{x}dy\Pi _{\rho }(y)-\int_{-\infty }^{x}dy\Pi _{\sigma }(y)\right)\right],
\end{eqnarray}
which represent fluctuations of CDW, SDW, SS and TS phases, respectively. The
correlation functions are defined as
\begin{equation}
R_i(x)=\langle :O_i(x)O_i^+(0):\rangle.
\end{equation}
After some cacluation, we find that the correlation functions behavior as
\begin{equation}
R_i(x)\sim |x|^{-2+\alpha_i},
\end{equation}
The exponents $\alpha _{i}$'s determine the divergence of the
corresponding phase. The expressions obtained for the $\alpha _{i}$
are
\begin{eqnarray}
&&\alpha _{\rm CDW}=2-K _c\nu^{c }-K _s\nu^s,\\
&&\alpha _{\rm SDW}=2(1+|\gamma|^s)-K _c \nu^c-\mu^s / K _s ,\\
&&\alpha _{\rm SS}=2(1+|\gamma|^{c})-\mu^{c}/K _{c}-K _{s}\nu^{s},\\
&&\alpha _{\rm TS}=2-\mu^{c}/K _{c}-\mu^{s}/K _{s}.
\end{eqnarray}
with
\begin{eqnarray}
\mu _{c}&=&\frac{v_c}{v_+ + v_-}\left[ 1+\frac{v_s^2}{v_+
v_-}\left(1-\frac{\delta v^2}{\delta v_2^2}\right)\right],\\
\mu _{s}&=&\frac{v_s}{v_+ + v_-}\left[ 1+\frac{v_c^2}{v_+
v_-}\left(1-\frac{\delta v^2}{\delta v_1^2}\right)\right],\\
\nu _{c}&=&\frac{v_c}{v_+ + v_-}\left[ 1+\frac{v_s^2}{v_+
v_-}\left(1-\frac{\delta v^2}{\delta v_1^2}\right)\right],\\
\nu _{s}&=&\frac{v_s}{v_+ + v_-}\left[ 1+-\frac{v_c^2}{v_+
v_-}\left(1-\frac{\delta v^2}{\delta v_2^2}\right)\right],\\
\gamma_{c}&=&\frac{\delta v}{v_+ + v_-}\left[ 1+\frac{\delta
v_2^2-\delta v^2}{v_+ v_-}\right],\\
\gamma_{s}&=&\frac{\delta v}{v_+ + v_-}\left[ 1+\frac{\delta
v_1^2-\delta v^2}{v_+ v_-}\right].
\end{eqnarray}%
The ground state is controlled and named by the most divergent
correlation function, i.e., with the largest $\alpha_i$.

\begin{figure}
\includegraphics[width=8cm]{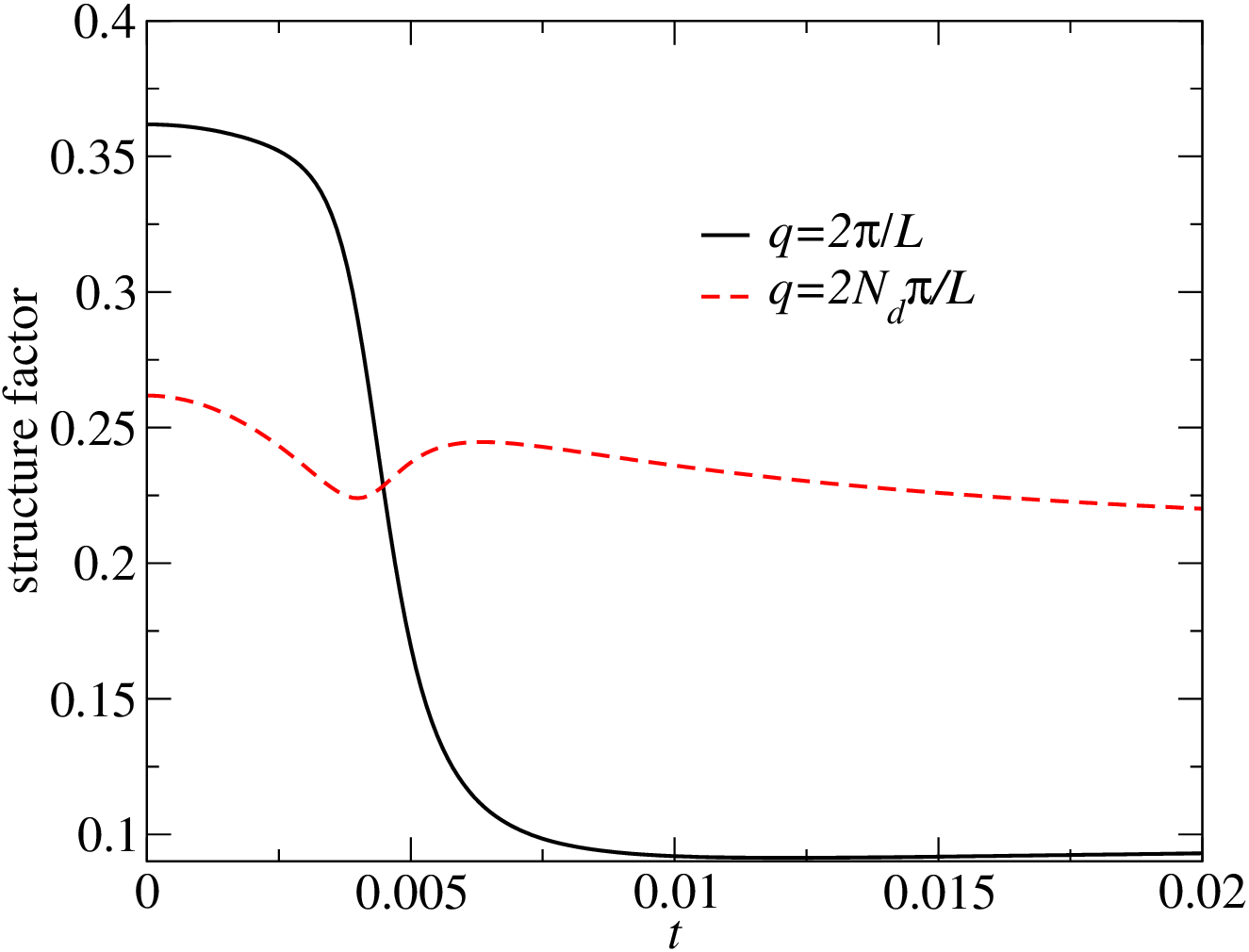}
\caption{\label{figure_cdw05} (color online) The structure factor of the CDW for
two different modes $q=\pi/L$ and $q=N_\downarrow \pi/L$ at given
$U/t_\uparrow=0.5$. Here $L=10, N_\uparrow=N_\downarrow=2,
t=t_\downarrow/t_\uparrow$, and $N_d$ denotes $N_\downarrow$.}
\end{figure}

\begin{figure}
\includegraphics[width=8cm]{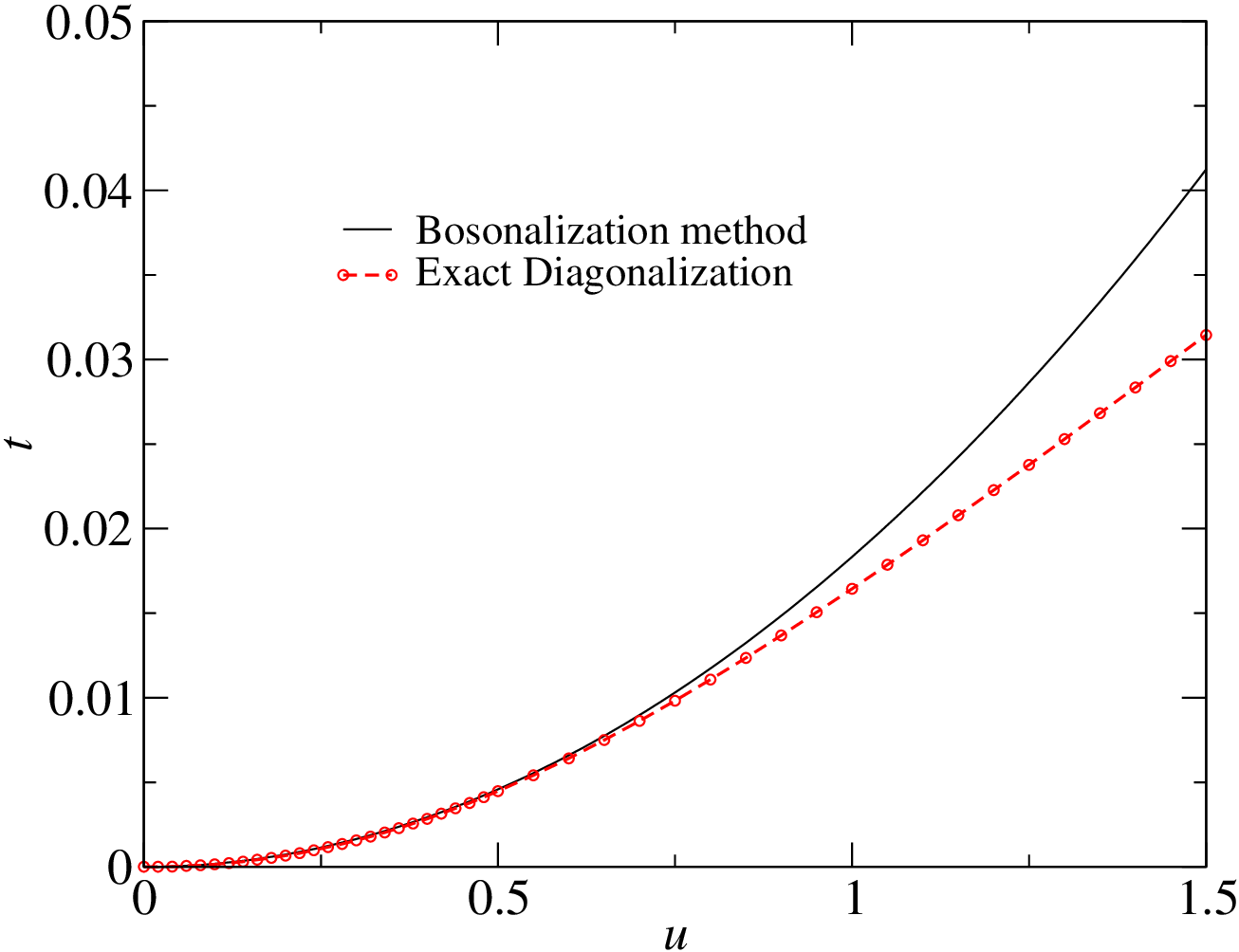}
\caption{\label{figure_tphase10_4} (color online) The boundary line between the
density wave and phase separation predicted by the bosonization method (solid
line) and exact diagonalization (dotted line for $N=10,
N_\uparrow=N_\downarrow=2$) at give concentration $n=2/5$. Here
$t=t_\downarrow/t_\uparrow, u=U/t_\uparrow$.}
\end{figure}

\begin{figure}
\includegraphics[width=8cm]{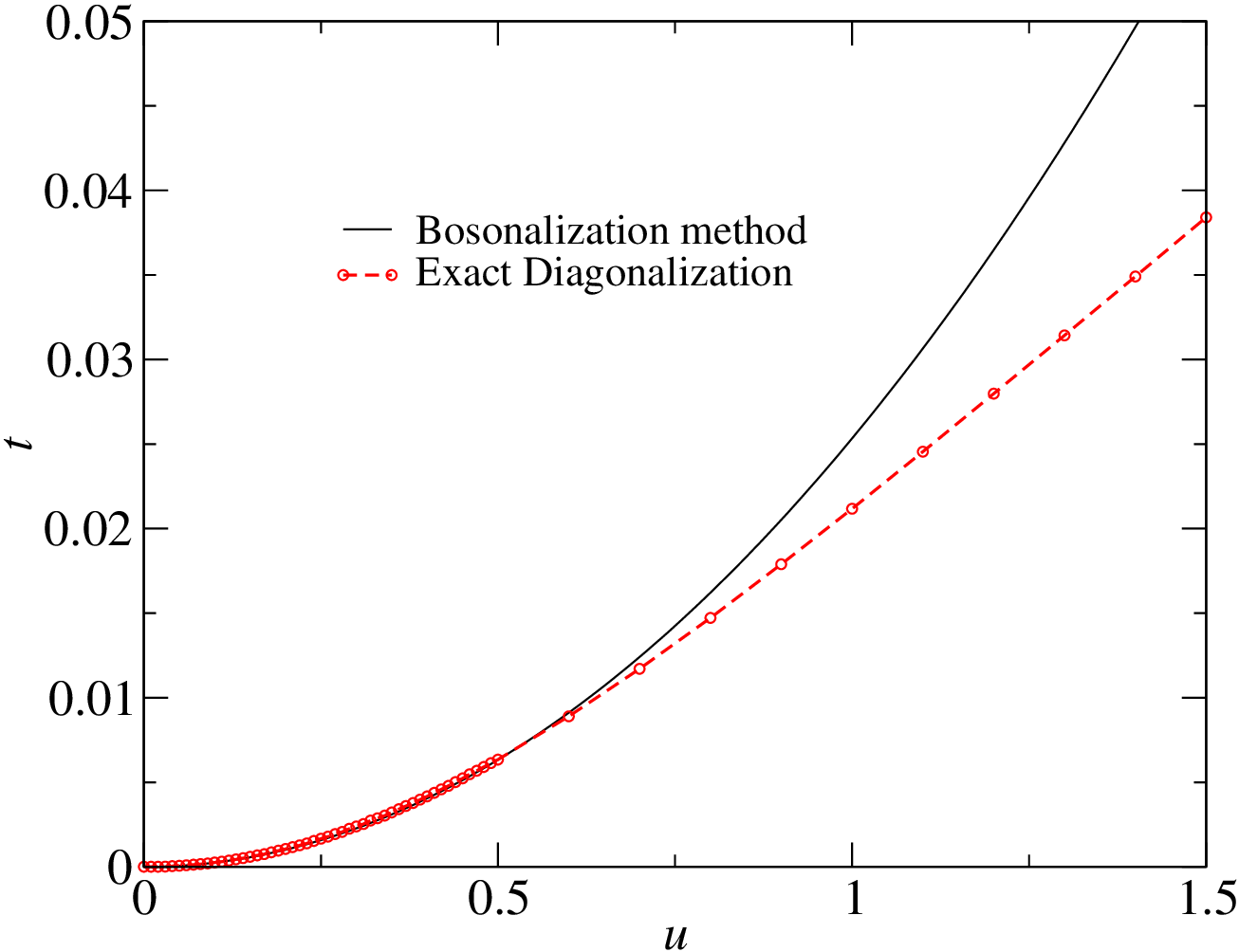}
\caption{\label{figure_tphase12_4} (color online) The boundary line between the
density wave and phase separation predicted by the bosonization method (solid
line) and exact diagonalization (dotted line, $N=12, N_\uparrow=N_\downarrow=2$)
at give concentration $n=1/3$. Here $t=t_\downarrow/t_\uparrow,
u=U/t_\uparrow$.}
\end{figure}

To have a deep understanding on the PS, it is very useful to study the structure
factor of the CDW. Since the dominating configuration of electrons with
spin-down is quite different in two phase, so we introduce the following
structure of down-spin electrons,
\begin{eqnarray}
S_{\rm CDW}(q) =\frac{1}{L}\sum_{jl}e^{iq(j-l)}\left(\langle n_{j,\downarrow}
n_{l,\downarrow} \rangle - \langle n_{\downarrow}\rangle^2\right),
\end{eqnarray}
where $q=2n\pi/L,\, n=0, 1, \cdots, L$. We show the structure fact as a function
of $t=t_\downarrow/t_\uparrow$ at a given $U/t_\uparrow=0.5$ for two different
modes $q=\pi/L$ and $q=N_\downarrow \pi/L$ in Fig. \ref{figure_cdw05}. From the
figure we find that in the small $t$ limit, $S(\pi/L)$ dominate, this fact
manifest the phase separation; while in a relatively larger $t$ region,
$S(N_\downarrow\pi/L)$ dominates, this suggests the density-wave state. So we
can use the intersection point of the structure factors of two different modes
to determine the transition point. In Fig. \ref{figure_tphase10_4} and
\ref{figure_tphase12_4}, we shown the phase diagram for different concentration
$n=2/5, 1/3$ in the small $U$ region. The results are very impressive. In both
figures, we can see that if $U < 0.5$, the numerical results from the exact
diagonalization method agree with Eq. (\ref{equ12}) excellently. That is the
phase boundary in the small $U$ regime is proportional to $U^2$. However, when
$U$ becomes large, say $U>1$, the bosonization results deviate from the
numerical results apparently. We interpret it due to the fact the bosonization
method becomes invalid in the large $U$ region. On the other hand, the excellent
agreement between the results obtained from two approaches suggests that
finite-size correction to the numerical data for a finite sample is very small.

Therefore, the bosonization results are wonderful in the small $U$ region and
low concentration conditions.  Since the 1D AHM is equivalent to the FKM if
$t_{\downarrow }=0$ $($or $t_{\uparrow }=0)$, which has been proven that there
exists PS phase at infinite-$U$ limit when the system shifts away form
half-filling. From Eq. (\ref{equ12}), we find that the PS phase always appears
in the 1D FKM whenever the onsite interaction is small or large as the system
shifts away form half-filling. On the other hand, the numerical studies
\cite{SJGu05} suggest there might exist a critical $U$ if the density of
electrons is close to half-filling. This inconsistence may due to the effect of
Umklapp process around half-filling.

\section{summary}
\label{sec:sum}

In summary, we have studied the quantum phase transitions in the 1D AHM with the
bosonization approach. In the framework of standard bosonization method, we
first obtained an effective Hamiltonian of 1D AHM. Then we diagonalized the
Hamiltonian and obtained the propagation velocities of the collective modes for
both spin and charge degree of freedom. Based on the instability condition, we
got the final critical conditions of the phase transition from DW to PS. We also
obtained the analytical expressions for the correlation functions of CDW, SDW,
SS, TS fluctuations, as well as the corresponding exponents.

Our results show that the difference between the hopping integrals for up- and
down-spin electrons is crucial for the happening of the PS. When the difference
is large enough, the phase separation will appear even if the on-site
interaction is small. In the small-$U$ and low concentration region, critical
conditions which scales like $t_\downarrow\propto U^2$ agree with the numerical
results excellently.

We thank HQ Lin for suggesting this problem to us. SJ Gu is grateful for the
hospitality of the Physics Department at Shanghai Jiaotong University. This work
is supported by RGC Grant CUHK 401504, NSFC Grant No. 10304014, and and
Foundations from Sci. \& Tech. Committee of Shanghai Municipality (Grant Nos.
03QA14054).


\end{document}